\documentclass[conference]{IEEEtran}
\IEEEoverridecommandlockouts
\usepackage{cite}
\usepackage{amsmath,amssymb,amsfonts}
\usepackage{algorithmic}
\usepackage{graphicx}
\usepackage{textcomp}
\usepackage{xcolor}
\def\BibTeX{{\rm B\kern-.05em{\sc i\kern-.025em b}\kern-.08em
    T\kern-.1667em\lower.7ex\hbox{E}\kern-.125emX}}
\begin{document}

\title{Magnetoelastics of High Field Phenomena in Antiferromagnets UO$_2$ and CeRhIn$_5$ \\
\thanks{Work at LANL and the National High Magnetic Field Laboratory was supported by the National Science Foundation through Cooperative Agreements No. DMR-1157490 and DMR-1644779, the State of Florida, and the U.S. Department of Energy, Office of Basic Energy Science, Div. of Mat. Sci. and Eng. MJ acknowledges support from LANL's Institute for Materials Science. KG acknowledges support from US DOE Early Career Research Program.  }
}

\author{\IEEEauthorblockN{1\textsuperscript{st}Marcelo Jaime}
\IEEEauthorblockA{\textit{National High Magnetic Field Laboratory} \\
\textit{and Institute for Materials Science}\\
\textit{Los Alamos National Laboratory}\\
Los Alamos, New Mexico 87545,  USA \\
mjaime@lanl.gov}
\and
\IEEEauthorblockN{2\textsuperscript{nd} Krzysztof Gofryk}
\IEEEauthorblockA{\textit{Idaho National Laboratory} \\
Idaho Falls, Idaho 83415, USA \\
krzysztof.gofryk@inl.gov}
\and
\IEEEauthorblockN{3\textsuperscript{rd} Eric D. Bauer}
\IEEEauthorblockA{\textit{MPA-CMMS} \\
\textit{Los Alamos National Laboratory}\\
Los Alamos, New Mexico 87545,  USA \\
edbauer@lanl.gov}

}

\maketitle

\begin{abstract}
We use a recently developed optical fiber Bragg grating technique, in continuous and pulsed magnetic fields in excess of 90T, to study magnetoelastic correlations in magnetic materials at cryogenic temperatures. Both insulating UO$_2$ and metallic CeRhIn$_5$ present antiferromagnetic ground states, at T$_N = $ 30.3K and T$_N =$ 3.85K respectively. Strong coupling of the magnetism to the crystal lattice degrees of freedom in UO$_2$ is found, revealing piezomagnetism as well as the dynamics of antiferromagnetic domain switching between spin arrangements connected by time reversal. The AFM domains become harder to switch as the temperature is reduced, reaching a record value H$_{pz}$(T = 4K) $\sim$ 18T. The effect of strong magnetic fields is also studied in CeRhIn$_5$, where an anomaly in the sample crystallographic $c$-axis of magnitude $\Delta$c/c $\simeq$ 2 ppm is found associated to a recently proposed electronic nematic state at H$_{en}\sim$ 30T applied 11$^o$ off the $c$-axis. Here we show that while this anomaly is absent when the magnetic field is applied 18$^o$ off the $a$-axis, strong magnetoelastic quantum oscillations attest to the high quality of the single crystal samples.

\end{abstract}


\section{Introduction}
The use of very high magnetic fields (H) for experimental studies of strongly correlated electron systems has proven to be a fertile area of condensed matter physics research in the last few decades. Part of the reason is that routine non-destructive magnetic fields produced in the laboratory can reach nowadays megagauss energy scales, namely around 100T. These fields, comparable to an energy scale of 10meV and a temperature scale of 100K, effectively compete with correlations existing between electronic, magnetic, and crystal lattice excitations in a large number of systems of current topical interest. The applied magnetic field is, hence, an external parameter that can be used to shift the balance between competing interactions and tune states that are otherwise difficult to observe \cite{jaime2000, jaime2002, zapf2014}, to overcome the effects of geometrical frustration in magnetic insulators \cite{jaime2012}, reversibly and continuously $focus$ experimental studies on phase transitions that are suppressed down to temperature (T) regimes where quantum fluctuations dominate over thermal fluctuations, alter electron/hole cyclotron orbits and facilitate fermiology studies \cite{harrisonXX} in real materials that exhibit a finite mean free path, and test predictions in topological matter such as massless Dirac fermions \cite{liu2017}, in addition to the traditional and highly valued mapping out of complex field-temperature (H,T) phase diagrams. Simultaneously, experimental probes for studies in high magnetic fields have reached a state where virtually all techniques that have once been used in any magnetic field at all are currently been developed or adapted for use in all magnetic fields available. Two prototypical examples of materials that have attracted the curiosity and intellect of the condensed matter physics community are UO$_2$ and CeRhIn$_5$.

\begin{figure}[h]
\centerline{\includegraphics[width=0.5\textwidth]{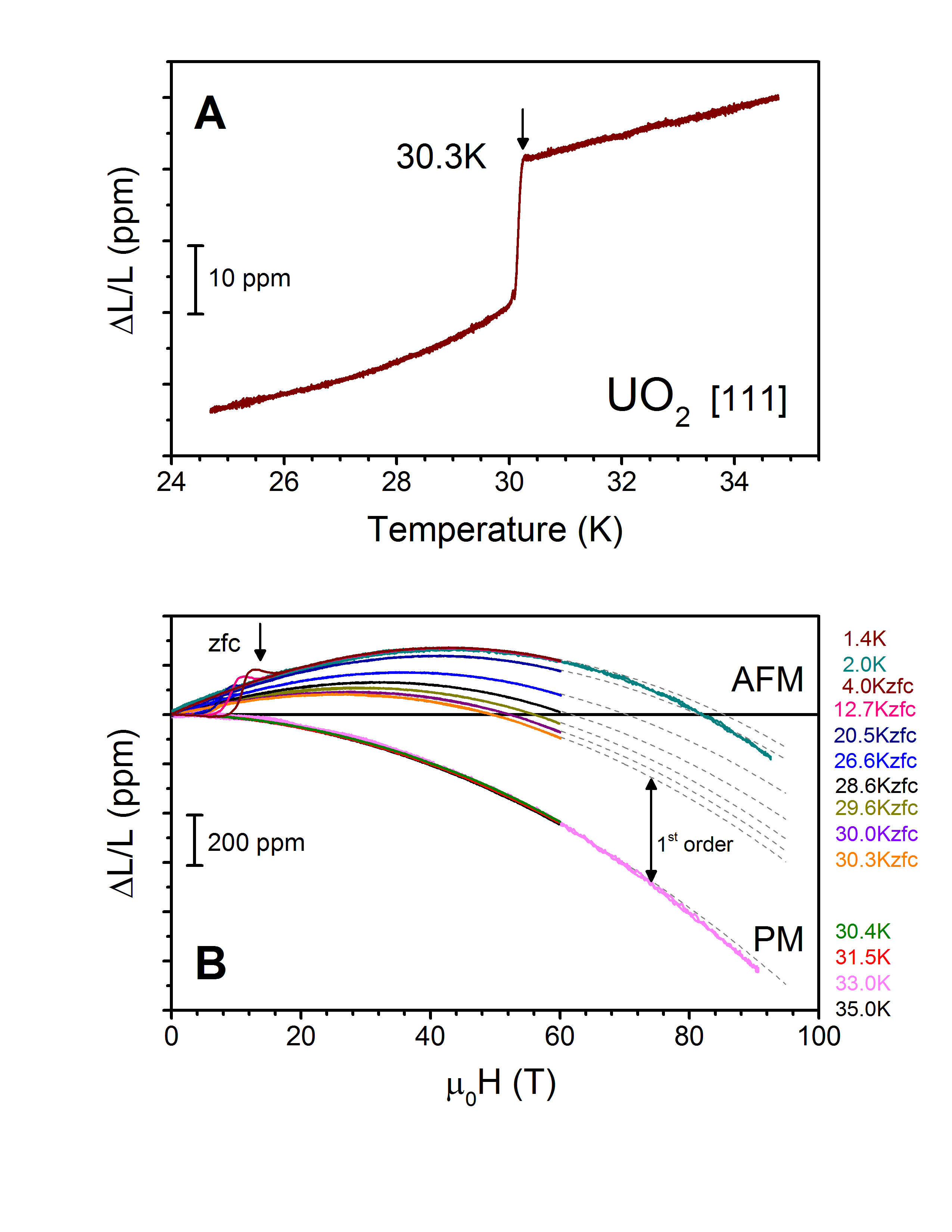}}
\caption{\bf{A}\rm. The thermal expansion of UO$_2$ $\Delta$L/L vs. temperature along the [111] crystallographic direction, measured in zero applied magnetic field. An abrupt drop reminiscent of a first order phase transition is seen at T$_N$=30.3K \bf{B}\rm. The axial magnetostriction $\Delta$L/L vs magnetic field applied along the [111] direction, measured at temperatures indicated between 1.4K and 35K. A linear term is evident in the otherwise simpler quadratic field dependence when the temperature is reduced below T$_N$. An anomaly is visible for fields under 20T on curves measured on zero field cooling (zfc) conditions, indicated by a vertical arrow. Dashed lines are fits with linear and quadratic terms on H.}
\label{fig1}
\end{figure}

UO$_2$ is a Mott insulator, known for more than seven decades to date and a core component of fuel cells used in power plants producing 20$\%$ of all human-made energy. It orders below T$_N$ = 30.3K in a 3k-type antiferromagnetic (AFM) structure where magnetic moments, at U-atoms occupying the nodes of an $fcc$ crystal lattice, point along the cubic body diagonals. Interestingly, the UO$_2$ solid resembles an organized collection of linear O-U-O molecules, and the orientation of magnetic moments in the 3k structure coincides with the orientation of their chemical bonds. The $fcc$ symmetry is, hence, preserved across the AFM first-order phase transition. Among the several outstanding puzzles presented by UO$_2$ stand its poor thermal properties, with a thermal conductivity $\kappa$(T=T$_N$) more than a hundred times smaller than isostructural non-magnetic ThO$_2$. The coincidence of this local minimum in $\kappa$(T) with the AFM ordering temperature is suggestive of a phonon mean free path that is dramatically reduced by some kind of magnetically correlated crystal bond disorder and/or soft-phonon phenomenon. The first order nature of the phase transition at T$_N$ makes this puzzle especially intriguing, and it has been proposed that magnetic fields could be a relevant tuning parameter \cite{gofryk2015, jaime2017}.

CeRhIn$_5$ is an anisotropic AFM metal that orders in a helix state below T$_N$ = 3.85K with a small ordered moment of 0.5~$\mu_B$ due to the Kondo effect and becomes a superconductor under an applied pressure p $>$ 1GPa, when the AFM ordering temperature is driven to zero. The AFM order, incommensurate helical along the tetragonal $c$-axis, can also be suppressed with external magnetic fields. Application of field in the basal plane suppresses the helix state at H$_M$ = 2.2T, giving place to a commensurate collinear square-wave AFM state. Surprisingly, magnetic fields in excess of 50T are required to completely remove the magnetic order, disregarding of the direction in which it is applied \cite{jiao2015}. While superconductivity does not survive such fields in CeRhIn$_5$, a puzzling Fermi surface reconstruction was observed when a field of H$_{en}\sim$30 T is applied along the crystallographic $c$-axis, accompanied by a rapid change in electrical properties \cite{moll2015}. Moreover, recent experiments show that the electrical properties become strongly anisotropic and correlated to the field orientation, pointing to a possible electronic nematic state \cite{ronning2017}. A few attempts have been made since to pinpoint this novel state \cite{rosa2018}, and more are needed to reach a satisfactory understanding.

Thermal expansion and magnetostriction are fundamental thermodynamic quantities that are directly derived from the Gibbs free energy G(p,T,H) of materials. Changes in G caused by varying external parameters such as pressure, temperature, or magnetic field can be studied by the thermal expansion coefficient $\alpha = \partial^2G/\partial{p}\partial{T}$ or the magnetostriction coefficient $\lambda = \partial^2G/\partial{p}\partial{H}$, respectively. Therefore, dilatometry techniques belong to the basic set of experimental probes present in materials science laboratories. These techniques are used alongside other fundamental magnetic, electric, and thermal capabilities to identify states of matter, to detect classical and quantum phase transitions between different ground states, and to understand the characteristics and nature of such transitions and transformations \cite{jaime2017b, rosa2017}. Here we use a recently developed optical fiber Bragg grating technique, in continuous and pulsed magnetic fields exceeding 90T, to study magnetoelastic correlations by means of high sensitivity dilatometry at cryogenic temperatures in UO$_2$ and CeRhIn$_5$ to better understand their zero field and field-induced low-temperature states.

\section{Experimental}

A single mode SiO$_2$ fiber Bragg grating technique is used for dilatometry studies of small single-crystalline samples of UO$_2$ and CeRhIn$_5$, which was developed for use in the extreme environments of very high, continuous, and pulsed magnetic fields of up to 150 T, at cryogenic temperatures down to 500mK in a $^3$He cryostat. In this technique Bragg gratings (FBG) are inscribed over a length of a 125$\mu$m optical fiber, normally tuned to reflect a particular wavelength of infrared light used in telecommunication. As strain sensors, the Bragg-reflected wavelength monitors the spacing of the grating and hence provides a measure of strain along the length of the sensitive region with $\Delta$L/L $\sim 10^{−8}$ precision. By attaching small FBGs to millimeter-size samples, one is able to detect small changes in sample length that are induced by magnetic fields (magnetostriction), or by changes in temperature (thermal expansion).

A commercial Hyperion\textsuperscript{\textregistered} swept wavelength laser system based on a tunable laser source was used in the 1500 - 1600 nm range to interrogate FBGs at frequencies up to 5 kHz, suitable for the sample environment in the bore of resistive magnets at the NHMFL DC Field Facility in Tallahassee, Florida. The swept wavelength nature of the light source in this type of application implies extremely low power (60 $\mu$W), making this instrument an ideal companion in cryogenic temperatures. In our 46 kHz interrogation scheme, suitable for sample environment in the bore of pulsed magnets at the NHMFL Pulsed Field Facility, the FBG is illuminated by a broadband white light source in similar infrared telecom spectrum using a commercial superluminescent light-emitting diode (SLED). A light polarization scrambler is used to minimize polarization rotation artifacts originated in the fringe field of pulsed magnets. The narrow spectral band around 1550 nm that is reflected by the FBG is diverted via a circulator to a 0.5 m spectrometer, where it is spectrally dispersed and detected by an InGaAs line scan camera. The strain sensitivity achieved with this approach is close to one part in 10 million (10$^{−7}$) \cite{jaime2017b}.

\section{Results and Discussion}

\subsection{UO$_2$}\label{UO2}

\begin{figure}[htbp]
\centerline{\includegraphics[width=0.5\textwidth]{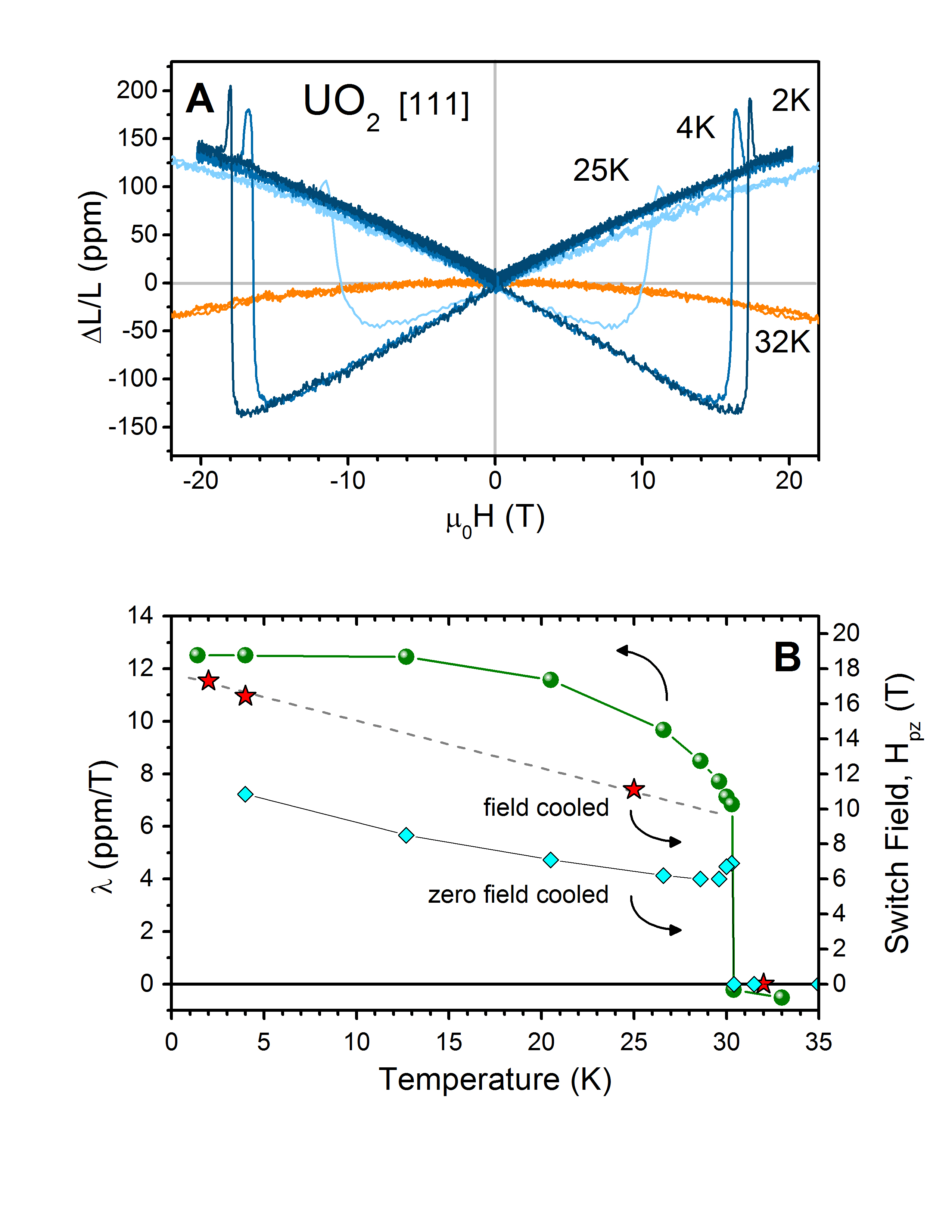}}
\caption{\bf{A}\rm. Axial magnetostriction $\Delta$L/L for fields along the [111] crystallographic direction, measured at different temperatures. A rapid increase at H$_{pz}$(T) indicates the abrupt flip of magnetic domains. No domain switch is observed in the paramagnetic state at T = 32K. \bf{B}\rm. Polynomial fits of the magnetostriction data in Fig.~\ref{fig1}\bf{B}\rm~ are used to extract the temperature dependence of the linear coefficient of magnetostriction $\lambda$(T) (green circles). The switching fields H$_{pz}$ defined at the local maximum of magnetostriction when sweeping the field in different directions (cyan rhombs are zfc, red stars are fc) increase with decreasing temperature.}
\label{fig2}
\end{figure}

The thermal expansion $\Delta$L/L(T) in zero mafieldgnetic , and axial magnetostriction $\Delta$L/L(H) in pulsed magnetic fields to 92.5T were measured at cryogenic temperatures on aligned single crystals of UO$_2$, cut along the $fcc$ body diagonal direction [111]. Fig.~\ref{fig1} shows the thermal expansion, where a clear anomaly is present at T$_N$, and magnetostriction data that follow a dominant quadratic field dependence in the high-temperature paramagnetic state. This simple behavior changes below the ordering temperature, where a linear-in-H term suddenly appears in magnetic fields in the 10-20 T range (see Fig.~\ref{fig1}\bf{A}\rm). Dashed grey lines show polynomial fits with only linear and quadratic terms. The first order nature of the AFM transition leads to an abrupt change of behavior at T$_N$ that is evident in the magnetostriction data. The linear term has been attributed to piezomagnetism in UO$_2$, a property characteristic of some non-collinear AFM systems where time-reversal symmetry is broken in a nontrivial way \cite{jaime2017}. Once the linear term is established by sweeping the magnetic field beyond the switching field H$_{pz}$, attributed to flipping of AFM domains connected by time reversal, subsequent magnetic field sweeps in opposite directions result in much sharper anomalies, as displayed in Fig.~\ref{fig2}\bf{A}\rm.

The linear-in-field coefficient displayed in Fig.~\ref{fig2}\bf{B}\rm~ (green circles) follows a temperature dependence that resembles an order parameter, reinforcing the notion of a piezomagnetism directly linked to the 3k AFM state. The magnetic point group symmetry of this state, [space group Pa$\overline{3}$, point group m$\overline{3}$] allows for the presence of a linear term in the magnetostriction \cite{jaime2017}. The axial magnetostriction observed, hence, amounts to experimental support for the known magnetic symmetry in UO$_2$. This is unusual for bulk probes, as scattering probes are usually required to test symmetry. The switching fields H$_{pz}$ (both zfc and fc) display an abrupt increase at T$_N$, followed by a linear increase as the temperature is further reduced. Magnetic field and temperature appear to cooperate in aligning pinned AFM domains, which are more easily flipped as T $\rightarrow$ T$_N$. The process, \it{i.e.}\rm~ the 'handle', by which AFM domains are flipped by a magnetic field in UO$_2$ remains so far unexplained. Also unexplained is the fact that extreme magnetic fields to 92.5T appear to be no match for the unusually robust AFM state, which remains unaffected to 60T even at fractions of a degree away from the ordering temperature.

\subsection{CeRhIn$_5$}\label{CeRhIn5}

\begin{figure}[htbp]
\centerline{\includegraphics[width=0.5\textwidth]{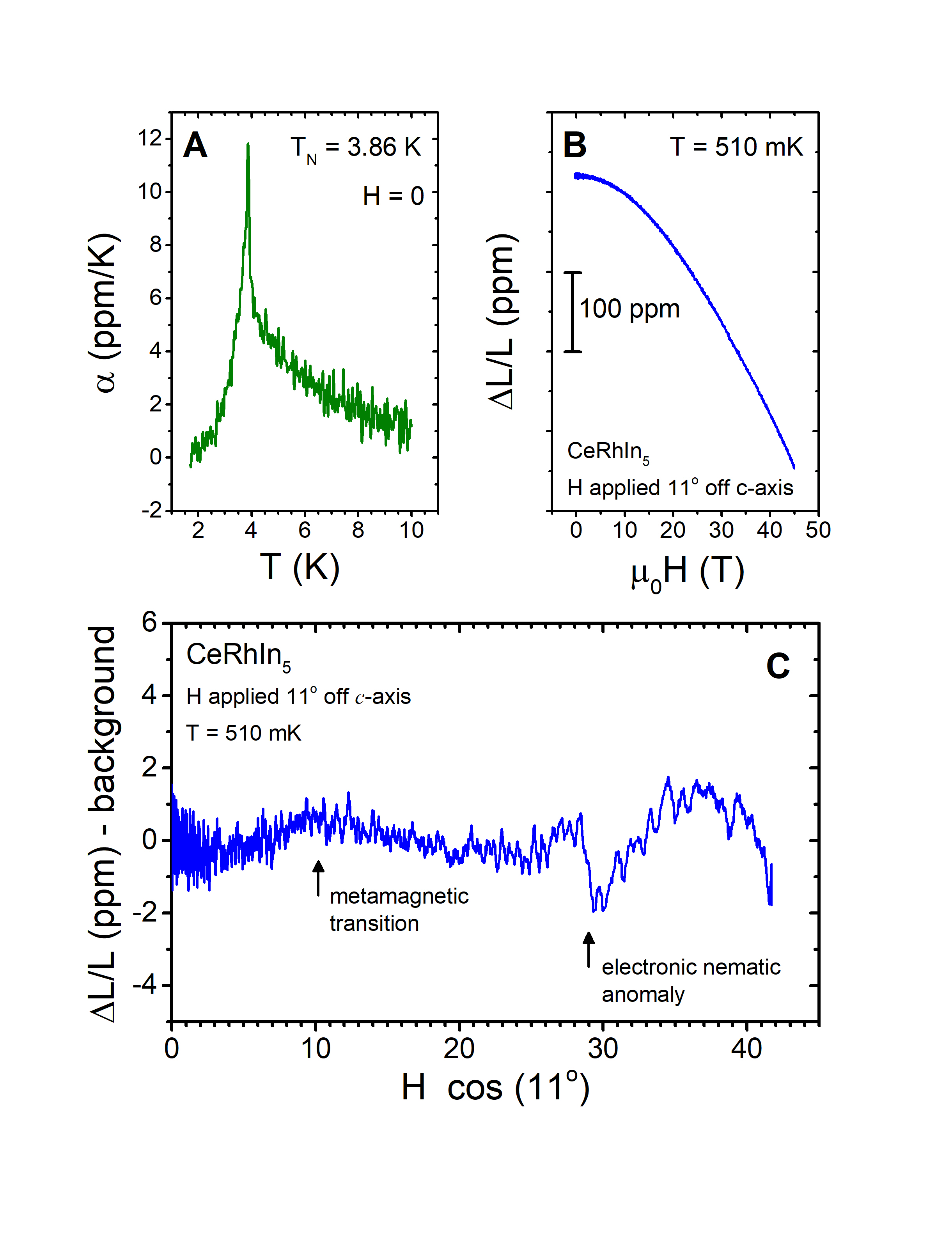}}
\caption{\bf{A}\rm. The linear coefficient of thermal expansion $\alpha$(T) for CeRhIn$_5$, measured along the tetragonal $c$-axis in zero field. The lambda-type anomaly at T$_N$ = 3.86K marks the onset to the low-temperature AFM state. \bf{B}\rm. Angle magnetostriction $\Delta$L/L // $c$-axis, with magnetic field applied 11$^o$ off the $c$-axis, towards a tetragonal principal axis measured at 510 mK. \bf{C}\rm. Angle magnetostriction after subtraction of a smooth polynomial background. Two anomalies are evident. One at lower fields corresponding to the helix to collinear order transition at 10-12T, another at 29T. The latter agrees well with a proposed electronic nematic state \cite{ronning2017}.}
\label{fig3}
\end{figure}

The thermal expansion of CeRhIn$_5$ was measured in zero applied magnetic fields at cryogenic temperatures. A sharp lambda-like anomaly in the coefficient of thermal expansion $\alpha$(T) (see Fig.~\ref{fig3}\bf{A}\rm) marks the onset of AFM order at T$_N$ = 3.86K, a hallmark of a second order type phase transition. The angle magnetostriction, $\Delta$L/L(H) measured in the AFM state along the crystallographic $c$-axis when the magnetic field is applied 11$^o$ off the $c$-axis, in pulsed fields to 45 T is shown in Fig.~\ref{fig3}\bf{B}\rm. Pronounced negative magnetostriction, quadratic in the field, is observed consistent with a magnetic point group that does not (as opposed to the case of UO$_2$ discussed above) break time-reversal symmetry in a nontrivial way. Subtraction of a polynomial background, however, reveals a rather interesting structure. A positive MS anomaly in the 10-12T range is likely associated with the low field helix to collinear AFM transition. A negative anomaly at H~$\times\cos$(11$^o$) = 29T agrees well with the proposed electronic nematic state \cite{ronning2017}. Only data collected during field down sweep is presented, as magnet/probe mechanical vibrations precluded clear observation of this anomaly during field up sweep. A full discussion of data taken in a continuous magnetic field to 45T, confirming this anomaly as well as quantum magnetoelastic oscillations only hinted here, is presented elsewhere \cite{rosa2018}.

\begin{figure}[htbp]
\centerline{\includegraphics[width=0.5\textwidth]{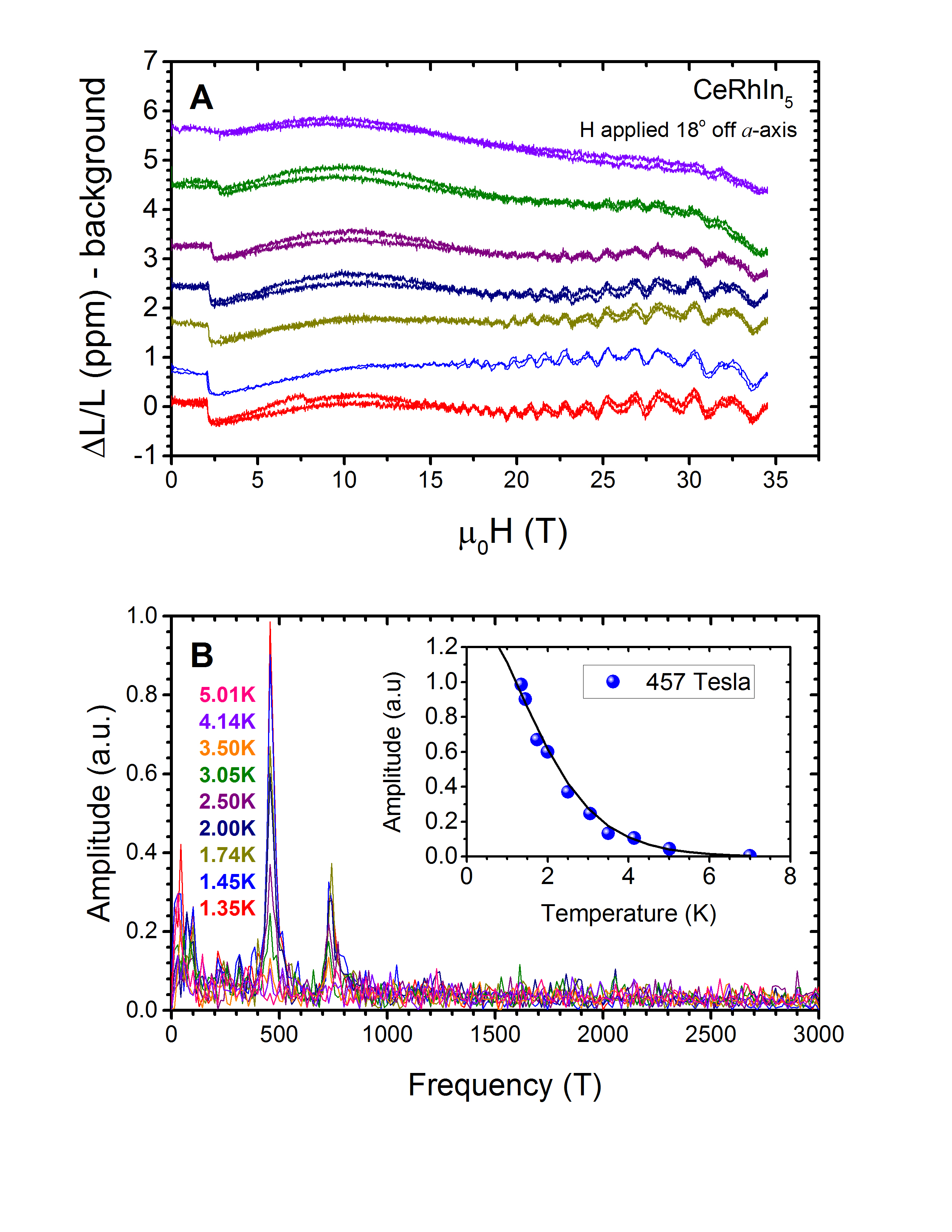}}
\caption{\bf{A}\rm. Angle magnetostriction $\Delta$L/L // $a$-axis for fields applied 18$^o$ off a tetragonal $a$-axis [100] in CeRhIn$_5$, after subtraction of a smooth polynomial background, measured at different temperatures. The anomaly at H$_M$ = 2.2T corresponds to the helix to collinear AFM structure transition. Quantum magnetoelastic oscillations are evident in the lowest temperature data sets. \bf{B}\rm. Fast Fourier transforms to the magnetostriction data where performed and plotted as amplitude vs frequency, for all measured temperatures. clear peaks are observed at various frequencies listed in Table \ref{tab1}. \bf{Inset}\rm. A Lifschitz-Kontsevich-type plot for the 457 T peak, showing good agreement for an effective electronic mass m$^*$ = 0.75m$_0$.}
\label{fig4}
\end{figure}

The Fermi surface effects uncovered by Jiao et al. \cite{jiao2015} and the electrical transport anomalies observed by Ronning et al. in CeRhIn$_5$ at H$_{en}$ are only present when the magnetic field is applied close to the tetragonal $c$-axis \cite{ronning2017}. In order to confirm that the magnetostriction anomaly is indeed associated to this novel high field electronic nematic state, we performed additional angle magnetostriction measurements with the magnetic field applied 17$^o$ off one of the tetragonal $a$-axis. Fig.~\ref{fig4}\bf{A}\rm~ displays the angle magnetostriction $\Delta$L/L // $a$-axis, when the magnetic field is applied 18$^o$ off the tetragonal $a$-axis, after subtraction of a smooth polynomial background, for various temperatures between 1.35K and 5.01K. While the magnetic transition at H$_M$ = 2.2T is sharp and clearly visible, the anomaly associated with H$_{en} \simeq$ 30T is completely lost. These results confirm our expectations. Meanwhile, the improved sensitivity of the FBG dilatometry technique in continuous magnetic fields reveal quite a nice quantum magnetoelastic oscillations in CeRhIn$_5$. These magnetoelastic oscillations, recently reported for fields close to the $c$-axis \cite{rosa2018}, are a serendipitous finding that attests to the high quality of the single crystal samples at hand.

\begin{table}[htbp]
\caption{FFT analysis of magnetoelastic oscilations}
\begin{center}
\begin{tabular}{|c|ccc|}
\hline
\hline

\textbf{FFT}&\multicolumn{3}{|c|}{\textbf{Frequencies (T)}} \\

\textbf{peak} & \textbf{ref. \cite{cornelius2000}$^*$}& \textbf{ref. \cite{hall2001}} &\textbf{\textit{This work}} \\

\hline

F$_1$& 105&  106 & 100 \\
F$_2$& 219& 212  & 214 \\
F$_3$& 295&  324 & 314 \\
F$_4$& 492&  446 & 457 \\
F$_5$& 714&  722 & 728  \\
F$_6$& 861& 874 & -  \\
\hline
\hline
\end{tabular}
\label{tab1}
\end{center}
\end{table}

Fig.~\ref{fig4}\bf{B}\rm~ shows the results of computing fast Fourier transformations (FFT) to the quantum magnetoelastic oscillations in CeRhIn$_5$. Peaks in the FFT are observed at frequencies that agree very well with de Hass-van Alphen oscillations \cite{cornelius2000,hall2001} (see Table~\ref{tab1}). Frequency changes due to magnetic fields applied 18$^o$ off [100] are not significant \cite{hall2001}. Our effective mass estimate for the FFT peak labeled F$_4$ is m$^*$ = 0.75 m$_0$, in agreement with Cornelius et al. Quite different are, however, the relative FFT amplitudes, indicating that electronic orbits involved in crystal bonding impacting quantum magnetoelastic oscillations are different than electronic orbits responsible for magnetism and impacting de Hass-van Alphen oscillations. One dramatic example is the almost complete suppression of F$_6$ in the quantum magnetoelastic oscillations reported here.

\section{Conclusions}

An optical fiber Bragg grating dilatometry technique was utilized to study oriented single crystal samples of insulating UO$_2$ and metallic CeRhIn$_5$, two very different antiferromagnetic systems, in the extreme sample environments of very high continuous and pulsed magnetic fields. A robust piezomagnetic state is explored in UO$_2$ and its switching fields H$_{pz}$, where the magnetic field applied along the $fcc$ [111] crystallographic direction induces flipping of AFM domains connected by time reversal, were mapped as a function of the temperature. We find that the field needed to switch AFM domains decreases with increasing temperature, and conclude that temperature and field must work together in this endeavor. These results show that some structure-sensitive probes, such as elastic neutron scattering where available practical magnetic fields are limited, could be utilized at temperatures close to T$_N$ = 30.3K where the required switch fields are in the $\sim$10T range. Such studies would be extraordinary tools to uncover the mechanisms behind the AFM domain pinning, as well as to explain the extraordinary resilience of the AFM state to external magnetic fields. The ability to bend the optical fibers used in the FBG dilatometry technique was exploited to study angle magnetostriction in CeRhIn$_5$ both along the tetragonal $a$- and $c$-axis. We find that an anomaly in the magnetostriction observed at H$_{en}$ when the magnetic field is applied close to the $c$-axis is absent when the field is applied close to the tetragonal $a$-axis. While the dilatometry measurements $\Delta$L/L are performed along specific axes, we know that all axes in a relatively simple structure such as tetragonal must be correlated by Poisson's rule. We then conclude that the anomaly found in the magnetostriction must be caused by the field-induced electronic nematic state. Moreover, we analyze quantum magnetoelastic oscillations for $\Delta$L/L // $a$-axis, H close to $a$-axis, and find excellent agreement with de Hass-van Alphen data in the literature attesting to the high quality of the samples at hand. These two seemingly unrelated systems, UO$_2$ and CeRhIn$_5$, serve as examples of how dilatometry studies can be of help to better understand underlying symmetries in magnetic systems. Plans for future work include the development of a modified dilatometer that allows for the simultaneous observation of two orthogonal crystallographic directions in the extreme environments of very high magnetic fields.

\section*{Acknowledgment}

We are indebted to our collaborators in this project M.B. Salamon, A. Saul, J. Lashley, J. Smith, T. Durakiewicz, A. Anderson, and C. Stanek. The most recent motivation to study high field dilatometry in CeRhIn$_5$ is owed to  P. Moll, while other contributors in the team include P.F.S Rosa, F. Balakirev, S. M. Thomas, J. Thompson, and F. Ronning.

\vspace{12pt}

\end{document}